\documentclass{eas}
\usepackage{graphicx}
\def\ltsima{$\; \buildrel < \over \sim \;$}    
\def\lesssim{\lower.5ex\hbox{\ltsima}}           
\def\gtsima{$\; \buildrel > \over \sim \;$}    
\def\gtrsim{\lower.5ex\hbox{\gtsima}}           

\newcommand{\BP}{Ballesteros-Paredes}
\newcommand{\gamef}{{\gamma_{\rm e}}}
\newcommand{\kms}{{\rm ~km~s}^{-1}}
\newcommand{\ls}{\lambda_{\rm s}}
\newcommand{\Ms}{M_{\rm s}}
\newcommand{\pcc}{{\rm ~cm}^{-3}}
\newcommand{\Peq}{P_{\rm eq}}

\newcommand{\tc}{\tau_{\rm c}}
\newcommand{\tturb}{\tau_{\rm turb}}
\newcommand{\VS}{V\'az\-quez-Se\-ma\-de\-ni}

\begin{document}

\title{Are There Phases in the ISM?} 
\author{Enrique V\'azquez-Semadeni}\address{Centro de Radioastronom\'\i
a y Astrof\'\i sica, Universidad Nacional Aut\'onoma de M\'exico, Campus
Morelia, P.O. Box 3-72, Xangari, Morelia, Michoac\'an, 58089, M\'exico.}
%
%
\begin{abstract}
The interstellar medium (ISM) is subject, on one hand, to heating and
cooling processes that tend to segregate it into distinct phases due to
thermal instability (TI), and on the other, to turbulence-driving
mechanisms that tend to produce strong nonlinear fluctuations in all the
thermodynamic variables. In this regime, large-scale turbulent
compressions in the stable warm neutral medium (WNM) dominate the
clump-formation process rather than the linear developent of TI. Cold
clumps formed by this mechanism are still often bounded by sharp density
and temperature discontinuities, which however are not contact
discontinuities as in the classical 2-phase model, but rather ``phase
transition fronts'', across which there is net mass and momentum flux
from the WNM into the clumps. The clumps grow mainly by accretion
through their boundaries, are in both thermal and ram pressure balance
with their surroundings, and are internally turbulent as well, thus also
having significant density fluctuations inside. The temperature and
density of the cold and warm gas around the phase transition fronts
fluctuate with time and location due to fluctuations in the turbulent
pressure. Moreover, shock-compressed diffuse unstable gas can remain in
the unstable regime for up to a few Myr before it undergoes a phase
transition to the cold phase, and is furthermore
constantly replenished by the turbulence. These processes populate the
classically forbidden density and temperature ranges. Since gas at all
temperatures appears to be present in bi- or tri-stable turbulence, we
conclude that the word ``phase'' applies only locally, surrounding phase
transition sites in the gas. Globally, the word
``phase'' must relax its meaning to simply denote a certain temperature
or density range.

\end{abstract}
\maketitle

\section{The standard multiphase models} \label{sec:std_multi}

\begin{figure}
\includegraphics[scale=0.35]{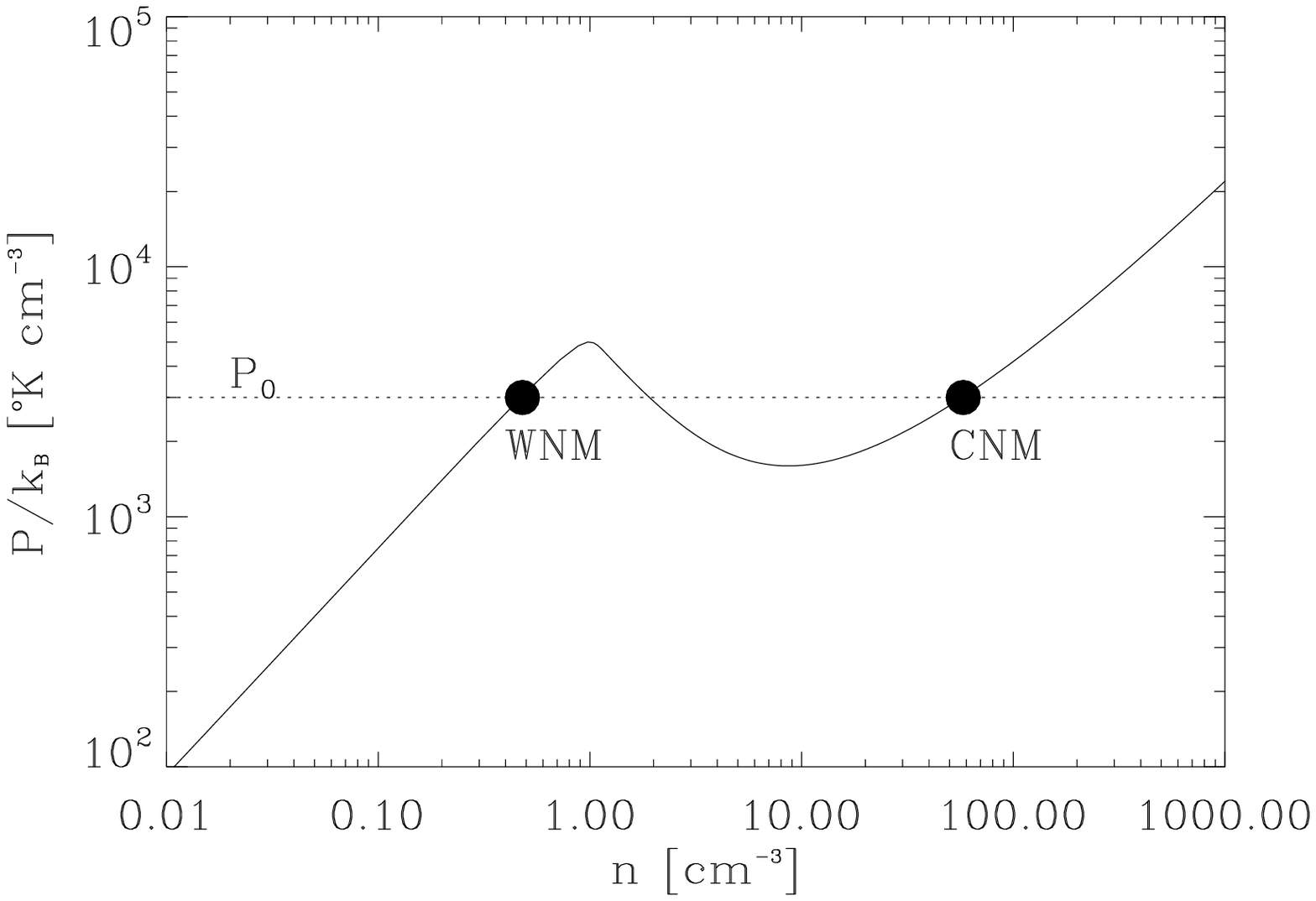}
\includegraphics[scale=0.35]{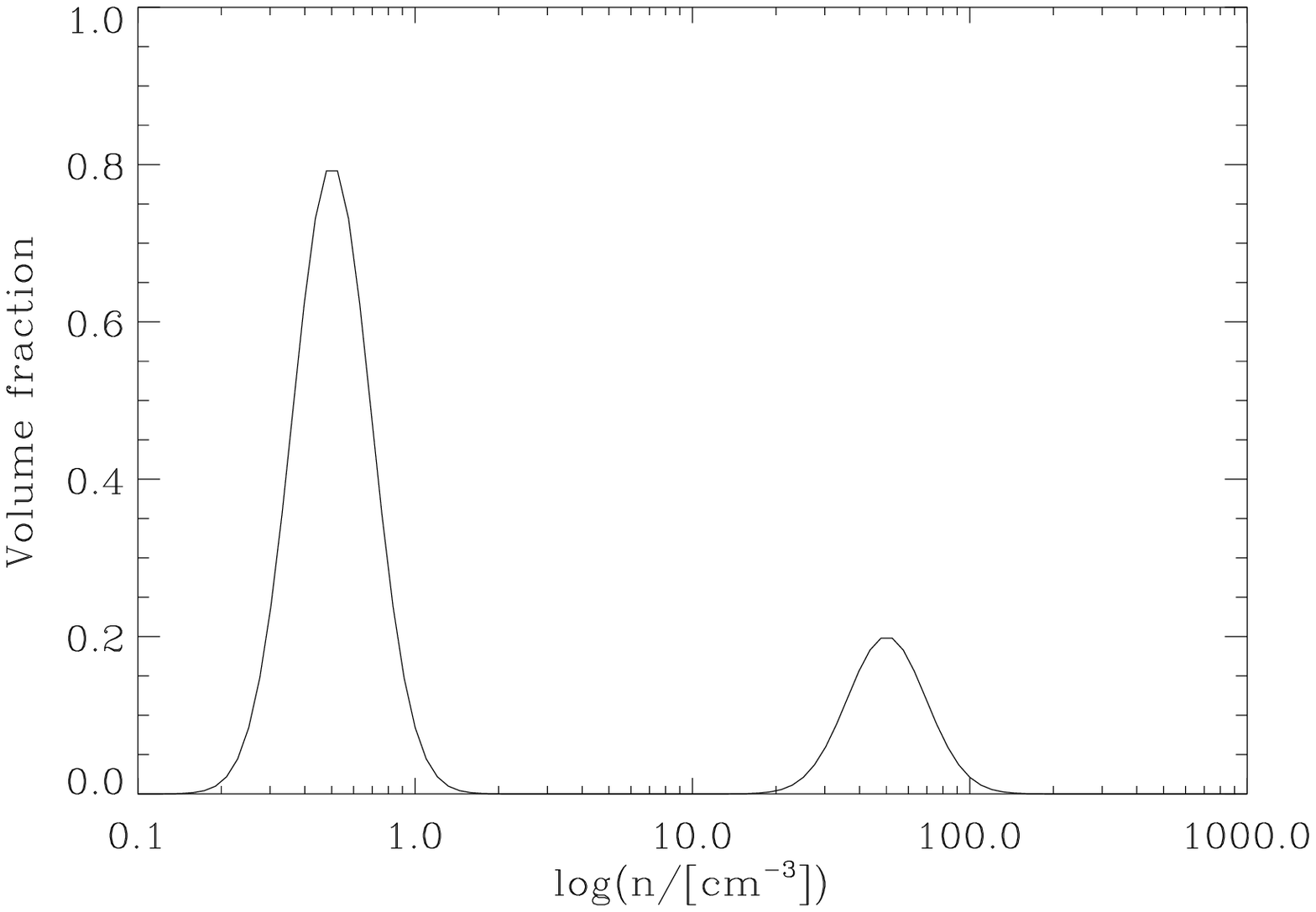}
\caption{{\it Left panel:} Thermal pressure corresponding to thermal
equilibrium between heating and cooling for the atomic ISM. Figure from
\cite{VS_etal07}, using the (errata-free) fit to the cooling function by
\cite{KI02}. The horizontal dotted line indicates a mean pressure $P_0$
that allows the medium to spontaneously segregate into a diffuse, warm
phase and a cold, dense one, indicated by the heavy dots. {\it Right
panel:} Schematic illustration of the density probability density
function (PDF) for the two-phase model. The vertical axis is in
arbitrary, non-normalized units, and the relative amplitude of the peaks
is meant to simply illustrate the fact that most of the volume is
occupied by the WNM.\label{fig:Peq}}
\end{figure}

The ISM is almost universally referred to as a ``multiphase'' medium, a
notion first proposed by \cite{Pikel68} and \cite{FGH69}. These
authors noted that, due to the net heating and cooling processes to
which the ISM is subject, the thermal pressure $\Peq$ corresponding to
thermal balance between heating and cooling is a non-monotonic function
of the density (fig. \ref{fig:Peq}, {\it left panel}). As is well known, the
negative-slope portion of this curve corresponds to a density range that
is unstable under the isobaric mode of the thermal instability (TI;
\cite{Field65}). 
Thus, if the ambient pressure lies within the
thermally unstable range, such as the pressure $P_0$ shown in Fig.\
\ref{fig:Peq} ({\it left panel}), the medium would tend to naturally
segregate into two stable {\it phases}\footnote{Thermodynamically, a
{\it phase} is region of space throughout which all physical properties
of a material are essentially uniform (e.g., Modell \& Reid, 1974). A
{\it phase transition} is a boundary that separates physically distinct
phases, which differ in most thermodynamic variables except one (often
the pressure).} at the same pressure, but very different densities and
temperatures, indicated by the heavy dots in the figure, and which are
commonly referred to as the warm and the cold neutral media
(respectively, WNM and CNM). This is the well-known ``two-phase'' model
of the ISM. In it, the final state of the instability would be a
collection of cold, dense clumps (the CNM) immersed in a warm, diffuse
intercloud medium (the WNM), in thermal and pressure equilibrium. The
density contrast between the cold clumps and the WNM was expected to be
$\sim 100$. The initial sizes of the cloudlets were expected to be
small,
%
and larger clouds
were assumed to form by merging (coagulation) of smaller cloudlets
(e.g., Oort 1954; Field \& Saslaw 1965; Kwan 1979). In this
scenario, the density and 
temperature probability density functions (PDFs, or histograms) of the
atomic ISM would consist of two Dirac delta functions at the values
corresponding to the WNM and the CNM, perhaps braodened by moderate stochastic
fluctuations in each phase (Fig.\ \ref{fig:Peq}, {\it right panel}). 

The two-phase model was later extended by McKee \& Ostriker (1977) to
include a third, hot phase, produced by supernova (SN) explosions,
leading to the famous three-phase model of the ISM. Supernova remnants
would start with large pressure excesses over the mean ISM thermal
pressure and expand until their pressures equalized with the ambient ISM
pressure. 

The two- and three-phase models were thus based on the notions of
simultaneous thermal balance between heating and cooling, and pressure
balance between the phases. Under such conditions, the clumps would be
bounded by contact discontinuities, since the pressure gradient between
the clumps and the WNM is zero. Interestingly, however, this notion was
maintained even in models in which the clumps were assumed to move
ballistically through the WNM.

\section{The controversy} \label{sec:controversy}

The classical discrete-phase models discussed in the previous section,
however, were not easy to reconcile with 
all observations. \cite{DST77} already noted that the range of spin
temperatures in HI gas is very broad, and in fact seemed to peak at
temperatures $\sim 300$ K, which correspond to the classically forbidden
unstable regime, $300 \lesssim T \lesssim 5000$ K. Since then,
significant amounts of gas in this temperature range have been
repeatedly found by various groups (e.g., Kalberla et al.\ 1985;
Spitzer \& Fitzpatrick 1995; Fitzpatrick \& Spitzer 1997; Heiles 2001a;
Kanekar et al.\ 2003), and moreover, a finer subdivision of the ISM
``phases'' has also been advocated on observational grounds (Heiles 2001b). 

On the numerical side, fully dynamic numerical simulations of the
turbulent ISM including parameterized heating and cooling, leading to a
``soft'' effective ``equation of state'' (EOS)\footnote{It can be shown
that the condition of thermal balance (heating = cooling), together with
the ideal-gas EOS, allows one to write an effective {\it polytropic} law
for the gas of the form $\Peq \propto \rho^{\gamef}$, where $\gamef$ is
the {\it effective polytropic exponent} (e.g., \VS\ et al. 1995, 1996),
and corresponds to the local slope of the curve in Fig.\
\ref{fig:Peq}. A more detailed perturbation analysis shows that the
actual value of $\gamef$ lies in between the thermal balance value and
the adiabatic value, depending on the ratio of the dynamical time to the
cooling time (e.g., Elmegreen 1991; Passot et al.\ 1995). An isothermal
flow corresponds to $\gamef=1$, while the isobaric mode of TI
corresponds to $\gamef <0$. Here, we adopt the convention that a
``soft'' EOS refers to $\gamef <1$. The gas is more compressible at
smaller $\gamef$, and in fact, as $\gamef
\Rightarrow 0$, the density jump due to a shock of Mach number $\Ms$
approaches $e^{\Ms^2}$ (\VS\ et al.\ 1996), exemplifying the large
compressibility of such regimes.} but no TI (Chiang \& Bregman
1988; Rosen \& Bregman 1995; \VS\ et al.\ 1995; Passot et al.\ 1995;
\BP\ et al. 1999a, 1999b), suggested that TI might not be an indispensible
channel to form cold, dense clouds in a diffuse, warm substrate, since
such a configuration was readily obtained in models with a soft EOS,
even if they were not thermally unstable.


Moreover, numerical
simulations at low resolution including TI (\VS\ et al. 2000), suggested
that turbulent mixing could smear the density PDF of the medium,
partially or completely erasing the multimodal signature of TI. A follow-up
numerical study at higher resolution (Gazol et al.\ 2001) showed that
the temperature PDF, although still bimodal, exhibited sizable amounts
of gas mass (up to $\sim 50$\%) in the unstable range between the CNM
and the WNM, in good agreement with observational estimates (e.g.,
Heiles 2001a; 
Heiles \& Troland 2003). More recent numerical models at high resolution
confirm these results (e.g., de Avillez \& Breitschwerdt 2005;
Hennebelle \& Audit 2007). These results all seemed to point towards a
continuum picture of the ISM (admittedly with large fluctuations in
density and temperature), rather than towards a medium composed of
various discrete phases. In fact, in an earlier study, Norman \& Ferrara
(1996), based on an analysis of the multiple scales at which turbulent
energy is injected into the ISM, went as far as proposing to extend the
notion of a multi-phase ISM to that of a ``phase continuum''. But this
proposal defies the very notion of well-differentiated, nearly uniform
{\it thermodynamic phases}.

\begin{figure}
\includegraphics[scale=0.37]{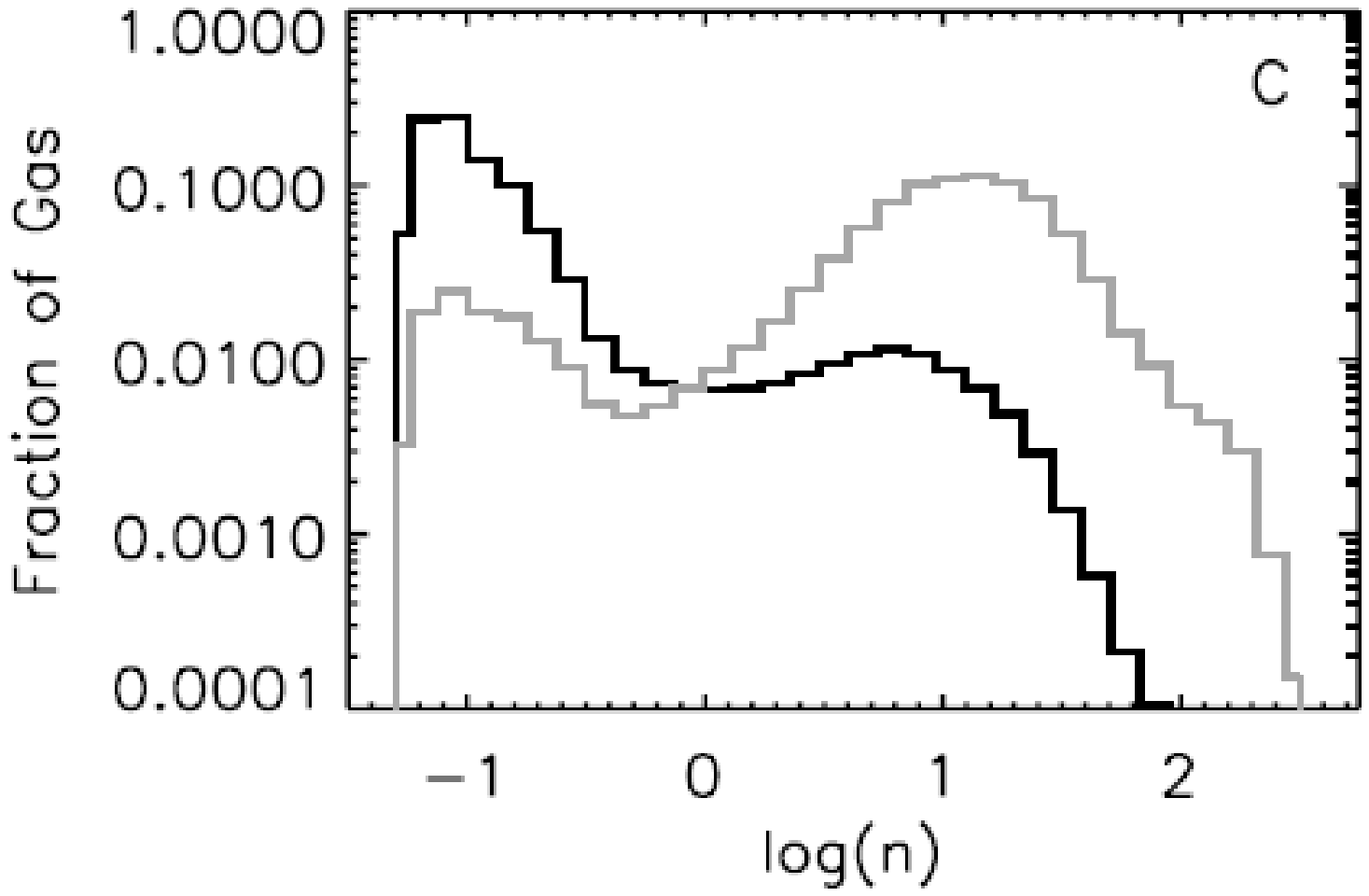}
\includegraphics[scale=0.3]{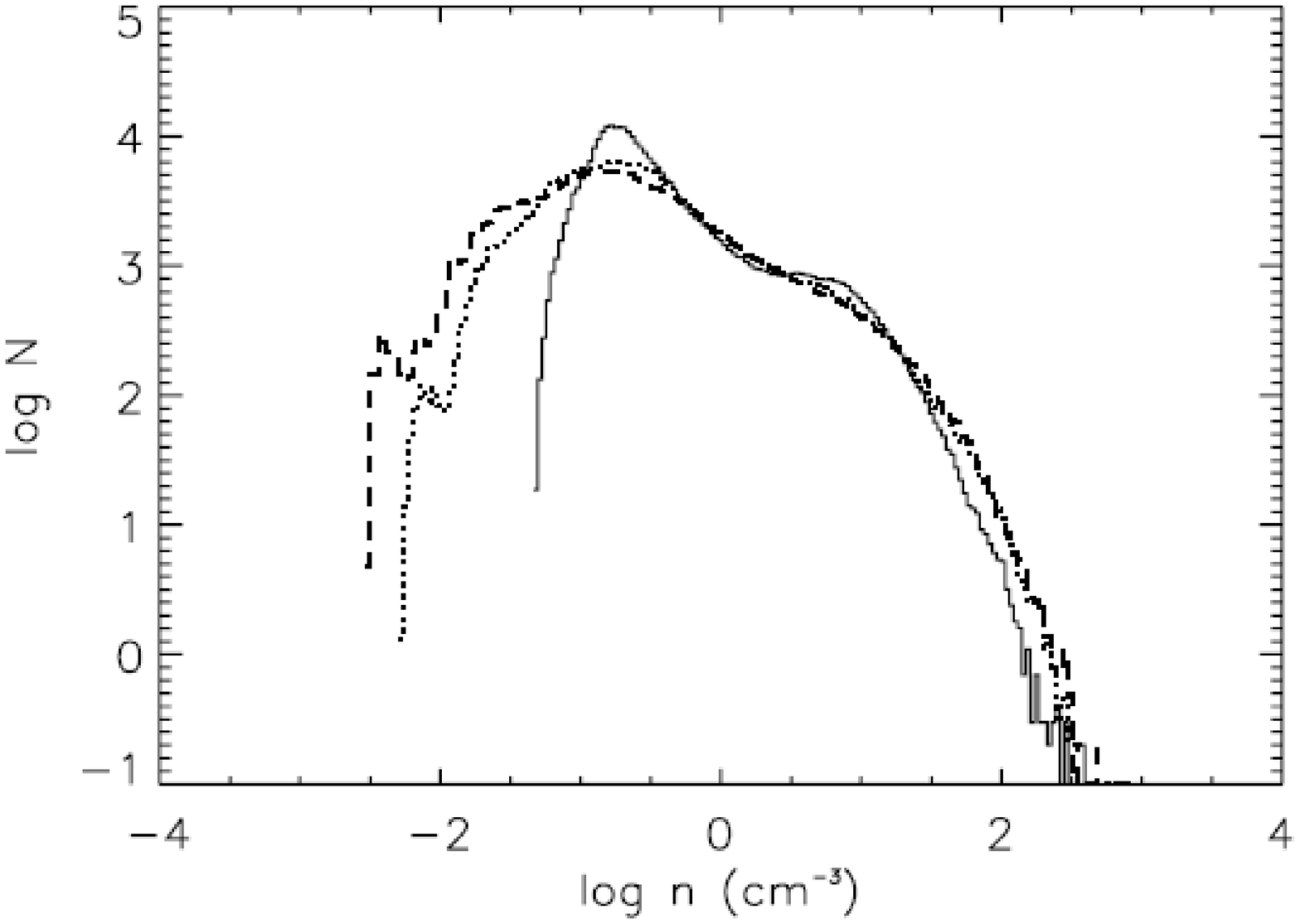}
\caption{{\it Left panel} Volume-weighted ({\it black line}) and
mass-weighted ({\it grey line}) density PDFs from a 3D numerical
simulation at $128^3$ resolution of the turbulent atomic ISM driven by
the magnetorotational instability, with a three-dimensional velocity
dispersion $\sim 2.7 \kms$ (from Piontek \& Ostriker 2005). {\it Right
panel:} Density PDFs of three 2D simulations with 2D velocity
dispersions $\sim 4.0 \kms$ ({\it solid line}), $\sim 9.1 \kms$ ({\it
dotted line}), and $\sim 11.4 \kms$ ({\it dashed line}) (from Gazol et
al. 2005).
\label{fig:ro_PDFs}}
\end{figure}

However, other numerical studies have advocated precisely the opposite,
i.e., that the two-phase model in particular continues to be applicable
for the turbulent atomic medium. For example, Piontek \& Ostriker (2005) have
found that, in three-dimensional (3D) low-resolution ($256^3$)
simulations of the atomic ISM driven by the magnetorotational
instability (MRI), even the most turbulent cases maintained the bimodal
shape of the density and temperature PDFs characteristic of the
two-phase model (fig.\ \ref{fig:ro_PDFs}, {\it
left frame}), while Audit \& Hennebelle (2005) and Hennebelle \& Audit
(2007) have pointed out that, in
two-dimensional (2D) simulations at very high resolution (up to
$10000^2$ grid points) of colliding
flows in the WNM, the classical two-phase picture still aplies in the
sense that the resulting CNM clouds and the WNM are still separated by stiff
thermal fronts, and the two media are in local pressure equilibrium.

\section{The hybrid regime of thermally bistable turbulence}
\label{sec:bistable_turb} 

The apparent controversy probably arises from the fact that the regime
of thermally bistable turbulence in the atomic ISM contains features of
both the classical, thermal- and pressure-balance two-phase regime, and
of a turbulent continuum, and in varying proportions depending on the
strength of the turbulence. In this section we first discuss the hybrid
nature of this medium, and then we extend the discussion to the full
range of interstellar densities and temperatures.


\subsection{Multimodality of the density and temperature PDFs}
\label{sec:multimodal_PDFs} 

Some of the most widely used statistical indicators of of compressible
turbulent flows are the PDFs of the density, temperature, and thermal
pressure. The transition from a discrete-phase PDF for a static,
pressure-balance two-phase model, such as that shown in Fig.\
\ref{fig:Peq} ({\it right panel}) to a unimodal PDF, such as that for a
single-phase polytropic flow (e.g., Scalo et al. 1998; Passot \& \VS\
1998; Nordlund \& Padoan 1999) appears to be continuous, rather than
abrupt. For example, Figure \ref{fig:ro_PDFs} ({\it right panel}) shows
density PDFs from three simulations by Gazol et al.\ (2005) with
increasing velocity dispersion. As the latter increases, the bimodality
of the PDF is seen to become less pronounced. This indicates the
increasing levels of turbulent mixing, which increasingly counteract the
segregation of the medium into two discrete phases.

The maintenance of a sizeable fraction of the gas mass in the thermally
unstable regime can be understood in terms of a competition between the
mixing and the cooling tendencies of the gas (see, e.g., the discussions
by Wolfire et al.\ 2003; \VS\ et al. 2003; Gazol et al. 2005). A
compressive turbulent velocity fluctuation is characterized by its
crossing time, $\tturb(L) \equiv L/v(L)$, where $L$ is the size scale of
the turbulent motions, and $v(L)$ is the turbulent velocity difference
between points separated by a distance $L$. If this time is shorter than
the cooling time of the gas, given by $\tc \approx c_V T/n\Lambda$,
where $c_V$ is the specific heat at constant volume, $n$ is the number
density, and $\Lambda$ is the cooling rate, the gas tends to respond
adiabatically to the compression, and is therefore thrown out of thermal
equilibrium between heating and cooling, becoming nonlinearly unstable
(Koyama \& Inutsuka 2000; Kritsuk \& Norman 2002a). Moreover, vortical
modes of the turbulence tend to mix the gas also in the turbulent crossing
time, fighting condensation.

The cooling time depends only on the local gas conditions, while the
turbulent crossing time is scale-dependent, typically following a
scaling law $v(L) \sim L^\alpha$, where $\alpha \sim 1/3 - 1/2$, the
former value being appropriate for incompressible (Kolmogorov, 1941)
turbulence, and the latter being appropriate for highly compressible,
Burgers-type (Burgers 1948) turbulence. Such a scaling implies that the
velocity dispersion is supersonic at large scales, and subsonic at small
scales, the two regimes being separated by the so-called {\it sonic
scale}, $\ls$. Based on observationally reported values of the global
velocity dispersion, and on estimates of the scales corresponding to
these values, Wolfire et al. (2003) estimated $\ls \sim 200$ pc for the
WNM. For this scale, they estimated a ratio of the characteristic times
$\Upsilon \equiv (\tc/\tturb)_{\lambda = \ls} \sim 1$ for the WNM,
suggesting that for this medium, a sizeable fraction of unstable gas
should exist, in agreement with numerical simulations with similar
parameters (Gazol et al. 2005; fig. 2, {\it right panel, dotted line}).

In general, it is then natural to expect that the fraction of gas in the
thermally unstable regime should increase for decreasing $\upsilon$
(i.e., higher turbulent rms Mach number), as indeed observed through
numerical simulations. This causes the density and temperature PDFs of
thermally bistable turbulence to transit smoothly from the stricly
bimodal shape characteristic of the two-phase model to a unimodal shape,
characteristic of a turbulent ``phase continuum'' (fig. 2, {\it right
panel}).

\begin{figure}
\includegraphics[scale=0.45]{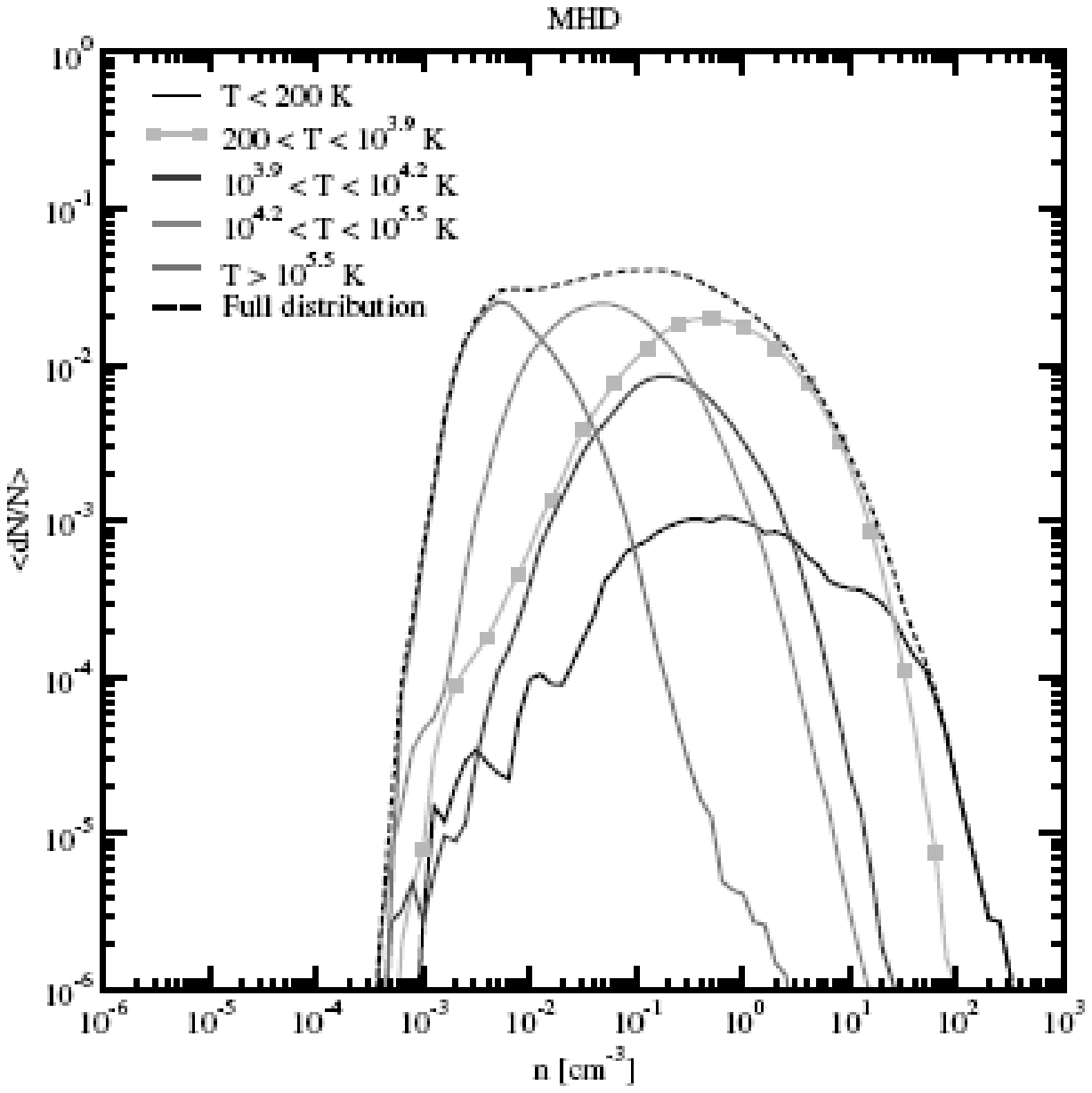}
\includegraphics[scale=0.45]{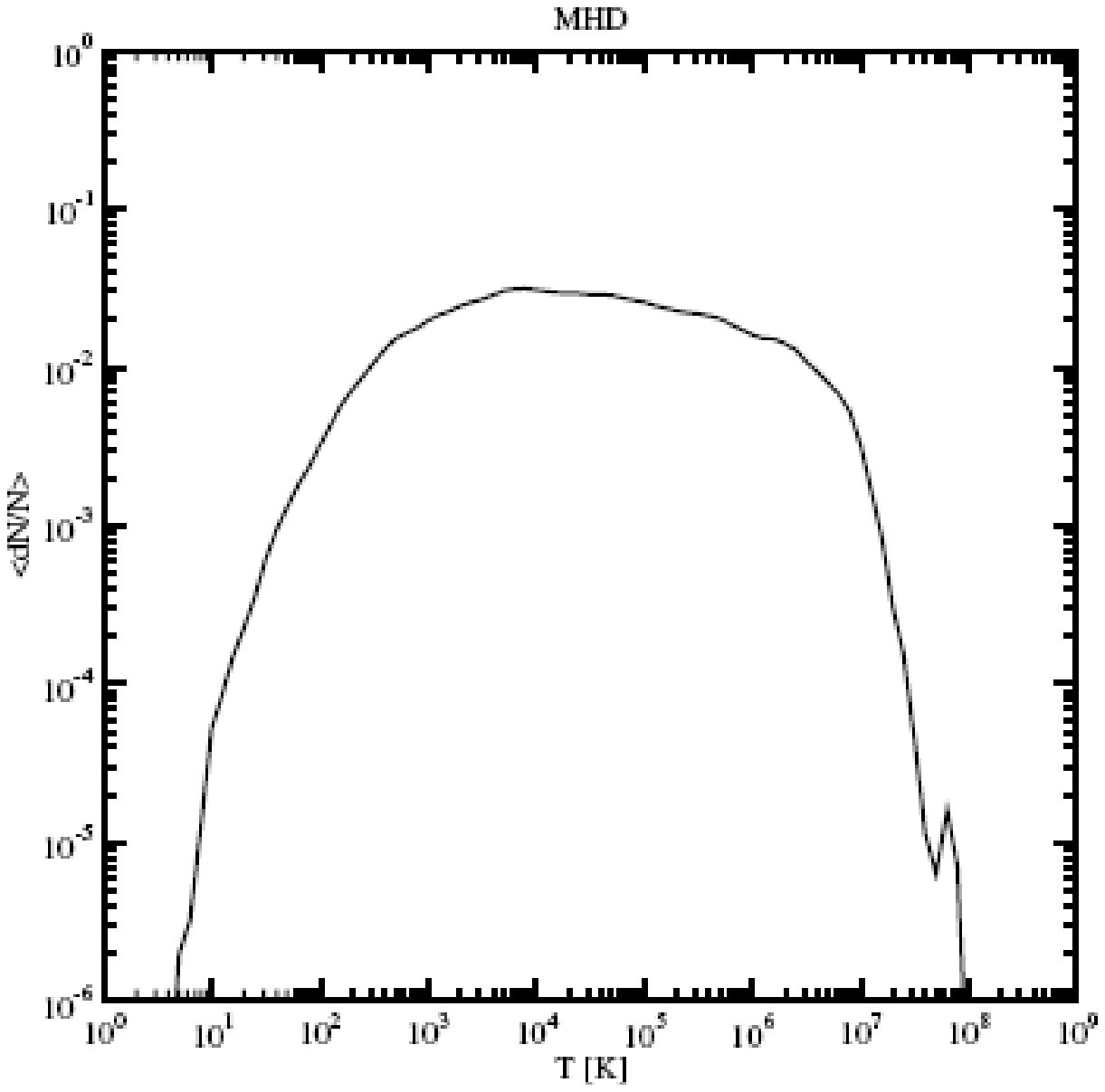}
\caption{Density ({\it left panel}) and temperature ({\it right panel}) PDFs
from numerical simulations of the SN-driven turbulent ISM by de Avillez
\& Breitschwerdt (2005). The density PDF shows a barely noticeable
multimodality, with subtle peaks at the ``hot-phase'' ($n \sim 10^{-2.5}
\pcc$) and ``warm-phase'' ($n\sim 10^{-1} \pcc$) densities. The signature of
the ``cold phase'' is barely noticeable as a kink at $n\sim 10^{2}
\pcc$. The temperature PDF has lost all signatures of
multimodality.\label{fig:dAB_PDFs}} 
\end{figure}

The above discussion can be extended to the three-phase model.
Numerical models of the ISM including SN driving of the turbulence (de
Avillez \& Breitschwerdt 2005), which include hot gas at temperatures up
to $T \sim 10^8$ K, exhibit density and temperature PDFs in which the
multi-modality is still present, albeit very mildly in the
hydrodynamical case, and barely noticeable in the MHD case. Figure
\ref{fig:dAB_PDFs} shows the corresponding PDFs for the MHD case, where
it can be seen that the density PDF ({\it left panel}) contains two very
subtle peaks at 
the densities of the ``hot'' and the ``warm'' phases (at $n \sim
10^{-2.5} \pcc$ and $n\sim 10^{-1} \pcc$, respectively). The signature of
the ``cold phase'' is barely noticeable as a kink at $n\sim 10^{2}
\pcc$. The temperature PDF ({\it right panel}) has lost all signatures
of multimodality, and 
only a relatively flat shape from $T \sim 10^2$ K to $T \sim 10^7$ K
remains, being indicative that the medium can exist at roughly
constant thermal pressure throughout this temperature range.

We conclude that in these SN-driven simulations, the density and
temperature PDFs are nearly flat over a wide
range, and the multimodal signature of truly distinct thermodynamic
phases has been nearly lost, with the unstable regimes being as
populated as the stable ones. In this case, the usage of the term
``multi-phase'' must be understood to mean ``multi-temperature'', and
the term ``phase'' to simply mean a range of temperatures or densities.

\subsection{Clump structure in thermally bistable turbulence}
\label{sec:clump_struct}

In view of the fact that the temperature PDFs of the turbulent ISM
become populated in the unstable ranges, the finding of Audit \&
Hennebelle (2005) that sharp discontinuities and thermal pressure
balance continue to hold at the boundaries of clumps is particularly
intriguing, and we now turn to how these two features can be reconciled.

A key issue in this problem is the fact that in a supersonically
turbulent medium, the compressive part of the velocity field works to
produce density enhancements (the ``clumps''; von Weizs\"acker 1951;
Sasao 1973; Elmegreen 1993; \BP\ et al. 1999a). Moreover, moderate
(transonic) turbulent compressions in the WNM (which is close to the
thermally unstable range) can nonlinearly trigger the development of TI
(Hennebelle \& P\'erault 1999; Koyama \& Inutsuka 2000, 2002), and thus
produce density enhancements of factors $\sim 100\times$ above the
WNM's mean density through compressions that would only cause
density enhancements of just a factor of a few in an isothermal flow.

\begin{figure}
\includegraphics[scale=0.4]{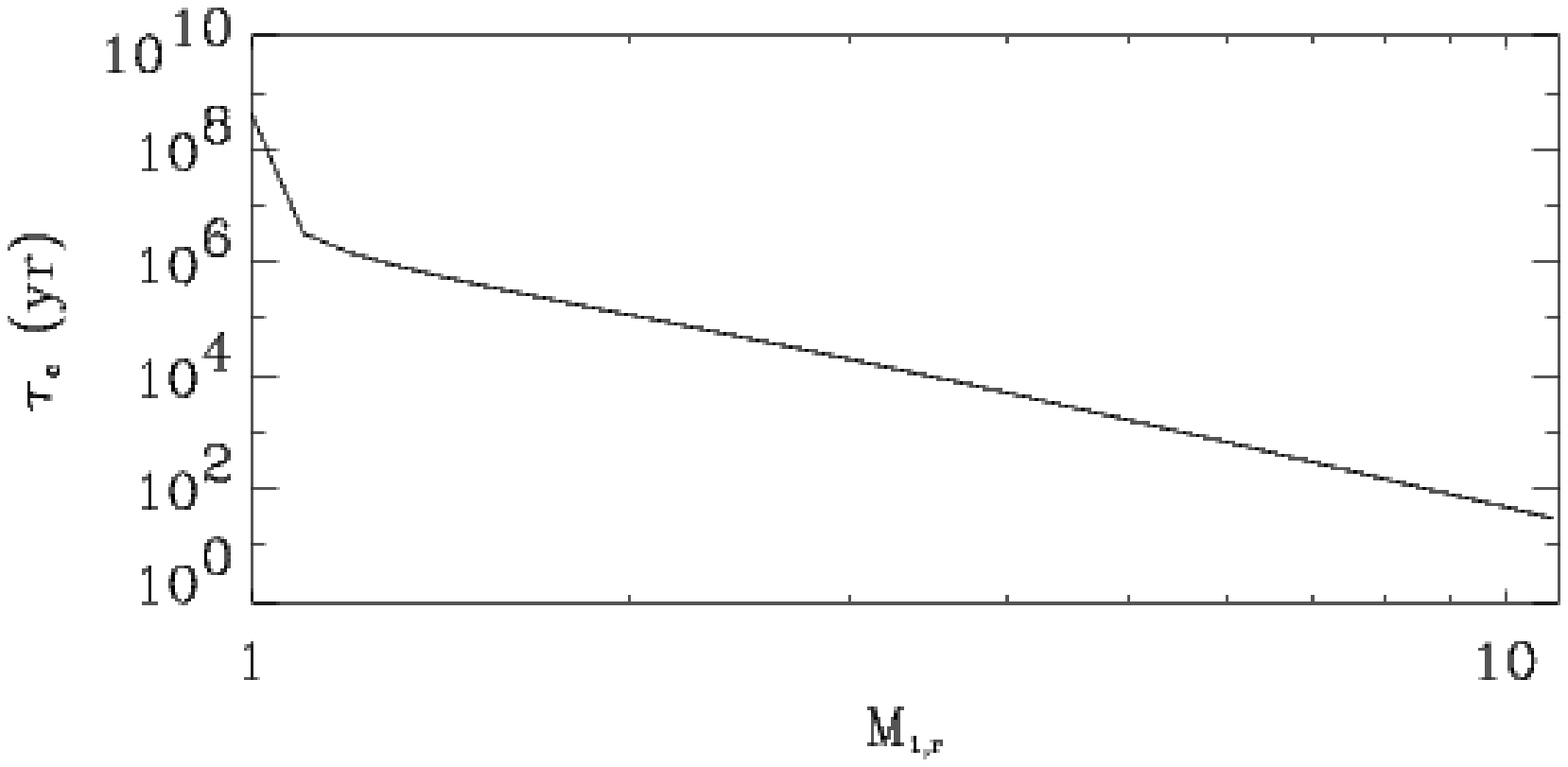}
\includegraphics[scale=0.35]{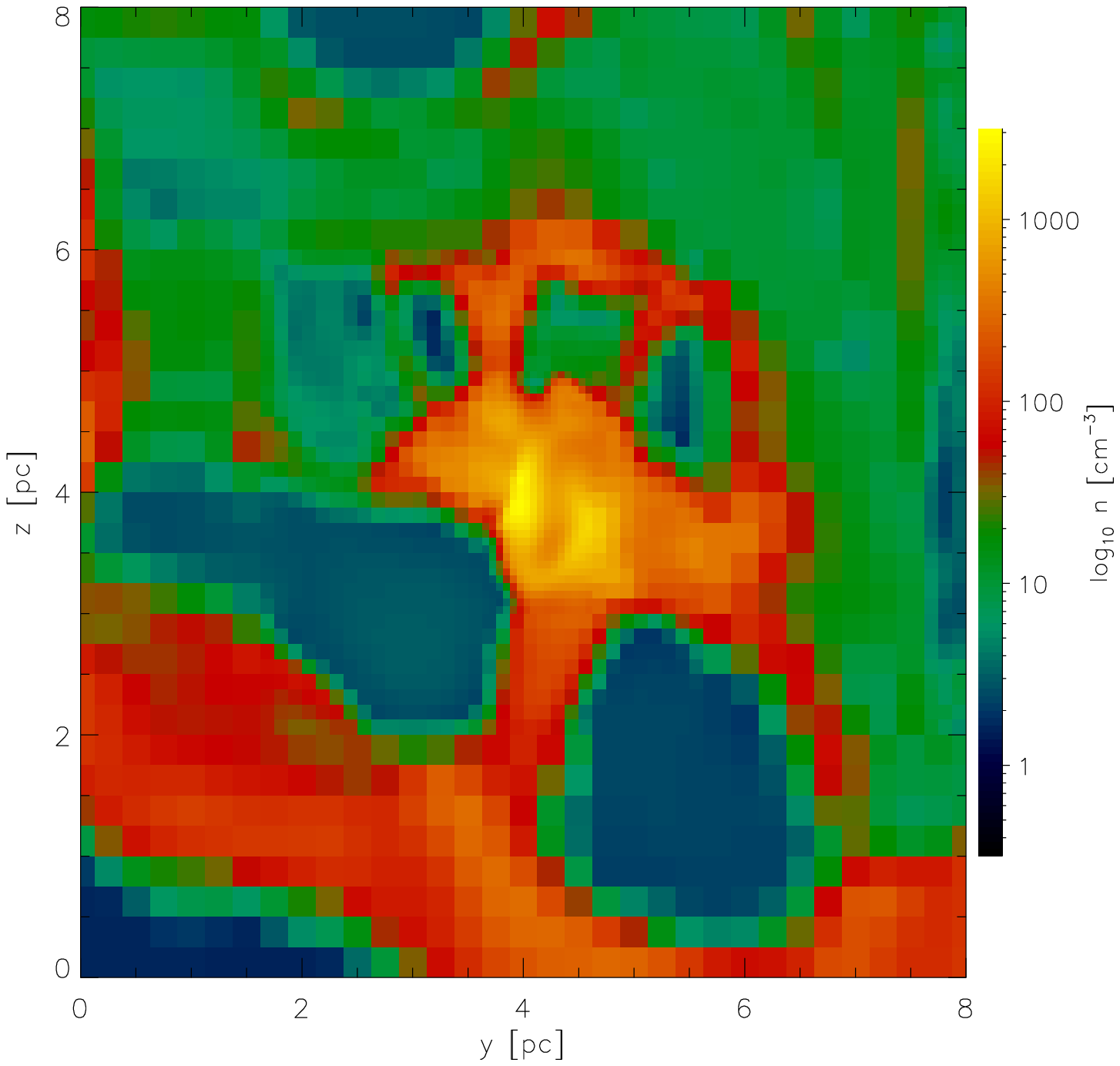}
\caption{{\it Left panel:} Cooling time $\tc$ for WNM gas after
having been shocked and compressed, as a function of the Mach number of
the compression (from \VS\ et al.\ 2006). After approximately $1 \tc$,
the shocked WNM undergoes a phase transition to the cold phase.
{\it Right panel:}Density cross section of a the central 8 parsecs of a
simulation by Banerjee et al.\ (2009). The full extent of the numerical
box is 256 pc. 
\label{fig:Robis_clump}}
\end{figure}

This process has been quantified by \VS\ et al. (2006), who presented
analytical solutions of the plane-parallel problem of a WNM flow
colliding against a wall (which represents the opposite colliding
stream) at transonic velocities, finding the physical conditions in the
resulting compressed layer of CNM as a function of the sonic Mach number
of the inflow. In this problem, an outer, outwardly-traveling shock is
formed, immediately behind which the (still warm) medium has been heated
and pressurized, and thus thrown away from cooling-heating
balance. Approximately one cooling time downstream from the shock
(fig. \ref{fig:Robis_clump}, {\it left panel}), the medium begins to be
able to cool, and eventually undergoes a {\it sudden} phase transition
to the CNM. This occurs at a ``phase transition'', or ``condensation''
front, which constitutes the boundary of the dense cloud. 

The main distinction between this process and the classical two-phase
model is twofold. First, there exists a continuous flux of mass across
the boundary, and thus the cloud is ``confined'' by the sum of the
thermal and ram pressures of the WNM inflow. We have placed quotes
around the word ``confined'' because the cloud is growing as it accretes
mass from, and is changing shape and sometimes being dispersed by, the
turbulent WNM surrounding it, rather than being truly confined in
place. Second, since the WNM has been heated by the shock, the process
is equivalent to one with a higher heating rate than the pure background
due to photoelectric heating from the diffuse UV background, an effect
whose net result is to shift the equilibrium pressure-density curve to
higher values of both variables (see, e.g., fig. 7 of Wolfire et
al. 1995). In a turbulent medium, in which there is a distribution of
shock strengths (e.g., Smith et al. 2000a,b), the equilibrium values of
the density and temperature of the WNM and CNM vary from one location to
another, thus populating classically forbidden density and temperature
ranges in the PDFs, while locally maintaining the two-phase structure
(Audit \& Hennebelle 2005; Hennebelle \& Audit 2007).


A recent study by Banerjee et al.\ (2009) has focused on the clump
structure within this framework, using the adaptive-mesh
refinement code FLASH (Fryxell et al. 2000) to simulate the formation of
dense 
clouds from transonic compressive motions in the WNM. The study includes
self-gravity and a weak (supercritical) magnetic field. They have found
that the clumps are indeed formed dynamically by the turbulent
compressions, being the densest regions in a network of sheets and
filaments. Figure \ref{fig:Robis_clump} ({\it right panel}) shows a
cross section of the central 8 parsecs of a simulation having a full
extent of 256 pc, and a mean field $\langle B \rangle = 1~\mu$G
permpendicular to the image. The density field is seen to exhibit sharp
jumps of up to factos of $\sim 100\times$, although a significant
fraction of the volume surrounding the clump is in the classically
unstable regime (colored {\it green} and {\it light blue}). In addition,
the central parts of the clump are seen to reach densities of up to $n
\sim 3000 \pcc$. This is $\sim 8$ times larger than the thermal
equilibrium density,\footnote{The simulation has a mean density of $1
\pcc$, uses a cooling curve leading to the $\Peq - n$ curve shown in
Fig.\ \ref{fig:Peq} ({\it left panel}) and starts with colliding streams
that have a Mach number (with respect to the WNM) of 1.22. With the
applied cooling curve, the temperatures of the warm and cold gas are
respectively $\sim 3000$ K and $\sim 20$ K. Thus, the implied
pressure-balance density of the cold gas, including the ram pressure of
the inflows, is $n \sim 375 \pcc$.} suggesting that turbulent motions
within the clump must be responsible for the additional density
enhancement by a factor of $\sim 8$.

The clumps in this regime are thus seen to exhibit features of both
turbulence-induced clumps, such as a centrally-peaked density profile,
amorphous shapes, and a turbulent internal velocity field, which induces
significant density fluctuations, while they retain features of the
two-phase model, such as sharp discontinuities at their
boundaries. These boundaries, however, are not contact discontinuities,
which by definition do not involve any mass flux across them, but rather
phase transition fronts, across which mass continually flows in from the
WNM, leading them to grow in mass and size, possibly being able to
eventually render them gravitationally unstable. The clumps are not
truly confined, since they move around, form part of larger filaments and
sheets, and grow in mass and size, being part of the global turbulent
flow, although they can be at much higher densities by virtue of the
thermal bistability of the flow.

Finally, it is worth noting that the simulations of Banerjee et al.\ (2009)
employed a very mildly supersonic inflow speed. It is quite likely that
higher inflow speeds may lead to a blurring of the sharp clump boundaries.

\section{Discussion and conclusions} \label{sec:conclusions}

So, are there phases in the turbulent ISM, then? The answer appears to
be more complicated than a simple ``yes'' or ``no''. At present, numerical and
analytic work suggests that, locally, yes, there exist clearly defined {\it
thermodynamic} cold and warm phases at each side of a phase transition
front but, 
because of the turbulent fluctuations in the pressure and the heating
rate, the densities and temperatures of these {\it locally stable}
phases
differ in general from one location to another, and moreover vary in
time as the local ram pressure varies, causing a constant flux of gas
among the phases, similarly to the effect of a temporal variation of the
background UV radiation field (Kritsuk \& Norman 2002b). Moreover, shocks
traversing the warm neutral gas leave it in an {\it unstable} regime,
from which it cools down to CNM conditions in roughly one cooling time,
which may amount to up to a few Myr, depending on the strength of the shock
(fig.\ \ref{fig:Robis_clump}, {\it left panel}). These processes
populate the entire range of values of density and temperature between those of
the classical cold and warm phases, creating a continuous
distribution. Although the details of the corresponding gap between the
classical warm and hot phases have not been investigated in the case of
a SN-driven turbulent ISM, it is conceivable that similar phenomena
occur, since the density and temperature PDFs in simulations of this
medium are equally densely populated. In this situation, no clear
dividing line exists between the hot, warm and cold gas, and the word
phase  is best understood as a range of density and temperature values,
while the true, distinct thermodynamic phases continue to exist locally
as a consequence of the accelerating nature of the cooling in thermally
unstable gas as it becomes denser, causing {\it sudden transitions} to
the cold phase.

Much work remains to be done in this field. In particular,
studies of the local structure of the gas similar to those that have
been performed for the atomic (warm and cold) ISM, are now necessary in
the context of the SN-driven ISM, in order to assess the nature of the
transition gas between the classical warm and hot phases.

\acknowledgements

The author is glad to acknowledge vivid and fruitful discussions
with Patrick Hennebelle over several years, and thanks him for a
critical reading of this manuscript. This research has been funded in
part by CONACYT grant U47366-F.



\begin{thebibliography}{99}

\bibitem[Audit \& Hennebelle (2005)]{AH05} Audit, E. \& Hennebelle,
P. 2004, A\&A 433, 1

\bibitem[\BP, Hartmann \& \VS (1999b)]{BHV99} Ballesteros-Paredes, J.,
Hartmann, L. \& V\'azquez-Semadeni, E. 1999b, ApJ 527, 285

\bibitem[\BP, \VS\ \& Scalo (1999a)]{BVS99} Ballesteros-Paredes, J.,
V\'azquez-Semadeni, E., \& Scalo, J. 1999a, ApJ, 515, 286

\bibitem[Banerjee et al.\ (2009)]{BVHK09} Banerjee, R., \VS, E.,
Hennebelle, P. \& Klessen, R.S. 2009, MNRAS, submitted (arXiv:0808.0986)

\bibitem[Burgers 1948]{Burgers48} Burgers,, J.M. 1948, Adv. in
Appl. Mech. 1, 171 

\bibitem[Chiang \& Bregman 1988]{CB88} Chiang, W.-H., \& Bregman,
J.N. 1988, ApJ, 328, 427

\bibitem[de Avillez \& Breitschwerdt 2005]{dAB05} 
de Avillez, M. A., \& Breitschwerdt, D. 2005, A\&A, 436, 585

\bibitem[Dickey et al. (1977)]{DST77} Dickey, J.M.,
Salpeter, E.E., \& Terzian, Y. 1977, ApJ, 211, L77

\bibitem[Elmegreen 1991]{Elm91} Elmegreen, B.G. 1991, in The Physics of
Star Formation and Early Stellar Evolution, ed. C.J. Lada and
N.D. Kylafis (Dordrecht:Kluwer), 35

\bibitem[Elmegreen (1993)]{Elm93} Elmegreen, B. G. 1993, ApJ 419, L29

\bibitem[Field 1965]{Field65} Field, G. B., 1965, ApJ 142, 531

\bibitem[Field et al.\ (1969)]{FGH69} Field, G. B.,
Goldsmith, D. W., \& Habing, H. J. 1969, ApJ, 155, L149

\bibitem[Field \& Saslaw 1965]{FS65} Field, G.B. \& Saslaw, W.C. 1965,
ApJ, 142, 568

\bibitem[Fitzpatrick \& Spitzer 1997]{FS97} 
Fitzpatrick, E. L. \& Spitzer, L. 1997, ApJ, 475, 623 

\bibitem[Fryxell et al. 2000]{FLASH00}
{Fryxell} B.,  {Olson} K.,  {Ricker} P.,  {Timmes} F.~X.,  {Zingale} M.,
  {Lamb} D.~Q.,  {MacNeice} P.,  {Rosner} R.,  {Truran} J.~W.,    {Tufo} H.,
  2000, ApJS, 131, 273

\bibitem[Gazol et al. 2001]{GVSS01}  Gazol, A., \VS, E.,
S\'anchez-Salcedo, F. J.\&  Scalo, J. M. 2001, ApJ, 557, 121 

\bibitem[Gazol, \VS\ \& Kim (2005)]{GVK05} Gazol, A., \VS, E. \& Kim,
J. 2005, ApJ, 630, 911

\bibitem[Heiles 2001a]{Heiles01a} Heiles, C. 2001, ApJ, 551, L105 

\bibitem[Heiles 2001b]{Heiles01b} Heiles, C. 2001, in Galactic
Structure, Stars and the Interstellar Medium, ASP Conference Series,
Vol. 231, ed. C. E. Woodward, M. D. Bicay, and
J. M. Shull (San Francisco: Astronomical Society of the
Pacific), 294 

\bibitem[Heiles \& Troland 2003]{ht03}
Heiles, C. \& Troland, T. 2003, ApJ, 586, 1067

\bibitem[Hennebelle \& P\'erault(1999)]{hen99} 
Hennebelle, P., \& P\'erault, M. 1999, A\&A, 351, 309

\bibitem[Hennebelle \& Audit (1999)]{HA07} 
Hennebelle, P., \& Audit, A\&A, 465, 431


\bibitem[Kalberla, Schwarz \& Goss 1985]{KSG85} Kalberla, P. M. W.,
Schwarz, U. J., Goss, W. M. 1985, A\&A, 144, 27

\bibitem[Kanekar et al 2003]{KSCS03}
Kanekar, N., Subrahmanyan, R., Chengalur, J. N., \& Safouris, V.
2003, MNRAS, 346, L57

\bibitem[Koyama \& Inutsuka (2000)]{KI00} Koyama, H. \& Inutsuka,
S.-I. 2000, ApJ, 532, 980

\bibitem[Koyama \& Inutsuka (2002)]{KI02} Koyama, H. \& Inutsuka,
S.-I. 2002, ApJ, 564, L97

\bibitem[Kritsuk \& Norman (2002)]{KN02} Kritsuk,, A., \& Norman,
M.L. 2002a, ApJ, 569, L127 

\bibitem[Kritsuk \& Norman (2002)]{KN02} Kritsuk,, A., \& Norman,
M.L. 2002b, ApJ, 580, L51

\bibitem[Kwan 1979]{Kwan79} Kwan, J. 1979, ApJ, 229, 567


\bibitem[McKee \& Ostriker (1977)]{MO77} McKee, C.F. \&
Ostriker, J.P. 1977, ApJ, 218, 148

\bibitem[Modell \& Reid (1974)]{MR74} Modell, M., \& Reid, R.C. 1974,
Thermodynamics and Its Applications (Englewood Cliffs, NJ:
Prentice-Hall) 

\bibitem[Nordlund \& Padoan 1999]{np99} 
Nordlund, \AA., \& Padoan, P.
1999, in Interstellar Turbulence, eds. J.\ Franco and A.\
Carrami\~nana (Cambridge: Cambridge University Press), p.\ 218

\bibitem[Norman \& Ferrara (1996)]{NF96} Norman, C.A. \& Ferrara, A. 1996,
ApJ, 467, 280 

\bibitem[Oort 1954]{Oort54} Oort, J.H. 1954,
Bull. Astr. Inst. Netherlands, 12, 177, no. 455 

\bibitem[Passot et al.\ 1995]{PVP95} Passot, T., \VS, E. \& Pouquet,
A. 1995, ApJ, 455, 536

\bibitem[Passot \& \VS\ 1998]{pv98} 
Passot, T., \& Vazquez-Semadeni, E. 1998, Phys. Rev. E 1999, 58, 4501

\bibitem[Pikel'ner (1968)]{Pikel68} Pikel'ner, S.B. 1968, Sov. Astron. 11,
737

\bibitem[Piontek \& Ostriker 2005]{PO05} Piontek, R.A. \& Ostriker,
E.C. 2005, ApJ, 629, 849

\bibitem[Rosen \& Bregman 1995]{RB95} Rosen, A. \& Bregman, J.N. 1995,
ApJ 440, 634

\bibitem[S\'anchez-Salcedo et al. 2002]{SVG02}
S\'anchez-Salcedo, F. J., V\'azquez-Semadeni, E., \& Gazol, A.
2002, ApJ, 577, 768

\bibitem[Sasao (1973)]{Sasao73} Sasao, T. 1973, PASJ, 25, 1

\bibitem[Scalo et al.\ (1998)]{SVCP98} Scalo, J., V\'azquez-Semadeni,
E., Chappell, D., \& Passot, T. 1998, ApJ, 504, 835 

\bibitem[Smith et al.(2000a)]{SMZ00a} Smith, M.D., Mac Low, M.-M., \&
Zuev, J.M. 2000a, A\&A, 356, 287

\bibitem[Smith et al.(2000a)]{SMZ00a} Smith, M.D., Mac Low, M.-M., \&
Heitsch, F. 2000b, A\&A, 362, 333

\bibitem[Spitzer \& Fitzpatrick 1995]{SF95} 
Spitzer, L. \& Fitzpatrick E. L. 1995, ApJ, 445, 196

\bibitem[V\'azquez-Semadeni et al.\ 2003]{VS_etal03}
V\'azquez-Semadeni, E., Gazol, A., S\'anchez-Salcedo, F. J., and Passot,
T. 2003, in Turbulence and Magnetic Fields in Astrophysics, ed. T.
Passot \& E. Falgarone, Lecture Notes in Physics, 614, 213

\bibitem[\VS\ et al. (2000)]{VS_etal00} \VS, E., Gazol, A., \& Scalo,
J. 2000, ApJ, 540, 271

\bibitem[\VS\ et al. (2007)]{VS_etal07} \VS, E., G\'omez, G. C.,
Jappsen, A. K., Ballesteros-Paredes, J., Gonz\'alez, R. F., \& Klessen,
R. S. 2007, ApJ, 657, 870

\bibitem[V\'azquez-Semadeni et al.\ (1995)]{vpp95}
V\'azquez-Semadeni E., Passot T., \&  Pouquet A., 1995, ApJ, 441, 702

\bibitem[V\'azquez-Semadeni et al.\ (1996)]{vpp96}
V\'azquez-Semadeni E., Passot T., \&  Pouquet A., 1996, ApJ, 473, 881

\bibitem[\VS\ et al. (2006)]{VS_etal06}
V\'azquez-Semadeni, E., Ryu, D., Passot, T., Gonz\'alez, R. F., \&
Gazol, A., 2006, ApJ, 643, 245

\bibitem[von Weizs\"acker (1951)]{von_Weiz51} von Weizs\"acker,
C.F. 1951, ApJ, 114, 165

\bibitem[Wolfire et al 1995]{Wol95} Wolfire, M.G.,Hollenbach, D.,
McKee, C.F., Tielens, A.G.G.M., Bakes, E.L.O. 1995, ApJ, 443, 152

\bibitem[Wolfire et al 2003]{wol03}
Wolfire, M.G., McKee, C.F.,  Hollenbach, D., \& Tielens, A.G.G.M.
2003, 587, 278

\end{thebibliography}
\end{document}